\documentclass[aps,onecolumn]{revtex4}
\usepackage{amsfonts,amssymb,amsmath,bm}
\usepackage[dvips]{graphicx}
\usepackage{color}

\begin{document}

\title{Nonlinear electro-hydrodynamics of liquid crystals.}

\author{E. S. Pikina$^{1,2}$, E. I. Kats$^1$, A. R. Muratov$^{2}$,  V. V. Lebedev$^{1,3}$.}

\affiliation{$^1$ Landau Institute for Theoretical Physics, RAS, \\
142432, Chernogolovka, Moscow region, Russia, \\
$^2$ Institute for Oil and Gas Research, RAS, 119917, Gubkina 3, Moscow, Russia,
$^3$ NRU Higher School of Economics, \\
101000, Myasnitskaya 20, Moscow, Russia, \\
}

\begin{abstract}

We present nonlinear  dynamic equations for nematic and smectic $A$ liquid crystals in the presence of an alternating electric field and explain their derivation in detail. The local electric field acting in any liquid-crystalline system is expressed as a sum of external electric field and the fields originating from feedback of liquid crystal order parameter, and a field, created by charged impurities. The system tends to decrease the total electric field, because it lowers the energy density. This basically nonlinear problem is not a pure academic interest. In the realm of liquid crystals and their applications, utilized nowadays modern experimental techniques have progressed to the point where even small deviations from the linear behavior can be observed and measured with a high accuracy. Hydrodynamics is the macroscopic description of condensed matter systems in the low frequency, long wavelength limit. Nonlinear hydrodynamic equations are well established to describe simple fluids. Similar approaches (with degrees of freedom related to the broken orientational or translational symmetry included) have been used also for liquid crystals. However to study behavior of strongly perturbed well above the thresholds of various electro-hydrodynamic instabilities of liquid crystals the nonlinear equations should include soft electromagnetic degrees of freedom as well. There are many examples of such instabilities, e.g., classical Carr-Helfrich instability triggered by the competitive electric and viscous torques, flexoelectric instability, and so one. Related to the occurring above the threshold states, manifest  a plethora of nonlinear phenomena, e.g., electro- or  magneto-optics, and electro-osmosis. Therefore the self-consistent derivation of the complete set of the nonlinear electro-hydrodynamic equations for liquid crystals became an actual task. The aim of our work is to present these equations, which is a mandatory step to handle any nonlinear phenomenon in liquid crystals.

\end{abstract}

\maketitle

\section{Preliminaries}
\label{sec:prel}

Nonlinear phenomena in general, and in liquid crystals particularly, are becoming one of the hottest topics in physics. Progress in experimental techniques and their accuracy has led to a number of new and exciting results still waiting for the theoretical analysis (or at least rationalization). To name a few such results, it is worth to noting three breakthrough publications \cite{lavrentovich1,lavrentovich2,lavrentovich3} (see also triggered by these works publications of other groups, where similar phenomena have been observed in different kinds of liquid crystals and  under different conditions \cite{dierking1,dierking2,dierking4,dierking5,dierking6,aya}). In these works, the authors presented the evidence of localized and propagating in a.c. electric field excitations in a liquid crystalline material. For these self-trapped wave excitations, exhibiting particle-like features system nonlinearity plays a key role, balancing excitations broadening (see, e.g., \cite{review1,review2} for Hamiltonian systems and \cite{akhmediev,TR16,QA22} for driven and damped (dissipative)
non-equilibrium systems). Specifically we focus in our manuscript on uniaxially symmetric types of liquid crystals, namely nematics ($N$) and smectic $A$ ($Sm A$) liquid crystals, which are ideal representative systems to study nonlinear phenomena.

Nematics \cite{GP93} are anisotropic fluids with a long-range orientational order defined by the unit headless vector, director $\bm n$. In smectics $A$ this uniaxial orientational anisotropy is supplemented by broken in one dimension translational symmetry. The starting point to investigate theoretically dynamic phenomena in liquid crystals is to derive the nonlinear dynamic equations. The complete set of the dynamic equations for any system should include all soft (hydrodynamic) variables caused both by general conservation laws and by symmetry breaking. For sufficiently low system conductivity $\sigma $, what is usually the case for liquid crystals, soft (slow for small $\sigma $) electromagnetic degrees of freedom also should be included in consideration. We did not find in the literature consistent derivation of  such general nonlinear electro-hydrodynamic equations for liquid crystals. The immediate motivation of our work is to present these equations which is a mandatory step to handle any nonlinear phenomenon in liquid crystals (see examples of problems, from about 50 years ago \cite{BJ73} and until very recent time, \cite{SS21}).

For the derivation of the equations we employ the universal Poisson brackets method \cite{DV80,KL93} permitting us to write out directly nonlinear non-dissipative dynamic equations automatically satisfying the conservation laws. Then the dissipative terms can be added based on the general thermodynamic relations providing positive entropy production, including besides orientational elastic energy, hydrodynamic motion, a.c. electric conductivity (nematics are typically weak electrolytes) and flexoelectric polarization. The obtained set of equations is not only of its own intellectual interest, but also of relevance to study many nonlinear phenomena in liquid crystals. To avoid too cumbersome expressions with too many unknown phenomenological parameters in this publication we restrict ourselves to only nematic and smectic $A$ liquid crystals. Thanks to their high symmetry (with an infinite order rotation axis) the equations can be represented in a relatively compact form. However if needed the equations can be formulated for less symmetric types of liquid crystals (e.g., for biaxial nematics or tilted smectic $C$ liquid crystals).

Here we derive the general hydrodynamic equations for a nematic or a  smectic-$A$ sample in an external electromagnetic field. The nonlinear dynamic equations of nematics and smectics-$A$ are known (see, e.g., \cite{BH88,DV80,KL93}. The linear theory of the low-frequency electro-hydrodynamic) instability also was developed long ago \cite{Pikin1991}.  However, a consistent nonlinear hydrodynamic theory with both reactive and dissipative (irreversible) terms, that includes soft degrees of freedom of the electromagnetic field as well,  is absent. This lacuna in the theory description exists  in spite of some insightful works \cite{PB96,PL04,LL10,TC16,PS17,PM18}. The matter is that in these publications the authors derived nonlinear dynamic equations for nematics or smectics $A$ by writing out all terms of a given order allowed by the symmetry, and then performing selection of the terms satisfying the conservation laws. The problem with such a method is that at each step taking into account next-order nonlinear terms, one has to repeat again the procedure and to check again that all conservation laws are satisfied.

The electromagnetic field in a continuous media is governed by Maxwell equations
\begin{eqnarray}
\partial_t \bm D
=c \nabla\times \bm H,
\label{maxw1} \\
\partial_t \bm B
=-c \nabla\times \bm E.
\label{maxw2}
\end{eqnarray}
Here $\bm E$ is the electric field, $\bm B$ is the magnetic field, $\bm D$ is the displacement field, $\bm H$ is the magnetizing field, and $c$ is speed of light. The equations (\ref{maxw1},\ref{maxw2}) have to be supplemented by the conditions for  divergencies of $\bm D$, $\bm B$. We assume
\begin{equation}
\nabla \bm D=0, \quad \nabla \bm B=0.
\label{maxw3}
\end{equation}
The relation $\nabla \bm B=0$ is exact and the relation $\nabla \bm D=0$ implies the electroneutrality of the medium, that is the charge density in the liquid crystal is assumed to be equal to zero. Both conditions (\ref{maxw3}) are, obviously, consistent with the dynamic equations (\ref{maxw1},\ref{maxw2}).

To close the system of equations (\ref{maxw1},\ref{maxw2}) one should relate $\bm D,\bm H$ to $\bm E, \bm B$. This deceptively simple exercise required a care in liquid crystals. To do this, the complete set of the nonlinear dynamic equations in the presence of the electromagnetic field is needed.

\section{Nonlinear Hydrodynamics of nematics}
\label{sec:hydro}

We start with the thermodynamic description of nematics based on thermodynamical identities. Considering the nematic in the first approximation as an ideal dielectric, one introduces the density of energy $U$, depending on density of mass $\rho$, density of entropy $S$, density of momentum $\bm J$, and director $\bm n$, as well as on the displacement field $\bm D_0$ and the magnetic field $\bm B$ \cite{LL84}. The notation $\bm D_0$ implies that one takes into account the contribution to the displacement field, related solely to polarization of the liquid crystal, as it should be in an ideal dielectric. The contribution to $\bm D$ related to conductivity (i.e., dissipation) has to be introduced separately, as a kinetic term.

The thermodynamic identity for the nematic in the presence of the electric and magnetic fields is a direct generalization of the thermodynamic identity for a solid dielectric, see Ref. \cite{LL84}, it is
\begin{equation}
dU=\frac{1}{4\pi}\bm E d\bm D_0+\frac{1}{4\pi}\bm H_0 d \bm B
+Td S+\mu d \rho + \bm v d \bm J
+\frac{\partial U}{\partial \bm n}d\bm n
+\frac{\partial U}{\partial (\partial_i\bm n)}d(\partial_i\bm n).
\label{wiakc4}
\end{equation}
Here $\bm E$ is the electric field, $\bm H_0$ is the magnetizing field, $T$ is temperature, $\bm v$ is velocity, and $\mu$ is chemical potential. As above for $\bm D_0$, the symbol $\bm H_0$ means that we consider solely the contributions to the magnetizing field related to the magnetic polarization, ignoring the kinetic processes. Note that according to Galilean invariance $\bm J=\rho \bm v$. Pressure $p$ is expressed as the Legendre transform
\begin{eqnarray}
p =\frac{1}{4\pi}(\bm D_0 \bm E+\bm B \bm H_0)+\rho\mu + S T+\bm J \bm v-U.
\label{pressure2}
\end{eqnarray}
One obtains from Eq. (\ref{pressure2}) the following thermodynamic identity
\begin{eqnarray}
dp =\frac{1}{4\pi}\bm D_0 d\bm E+\frac{1}{4\pi}\bm B d \bm H_0
+Sd T+ \rho d\mu + \bm J d \bm v
-\frac{\partial U}{\partial \bm n}d\bm n
-\frac{\partial U}{\partial (\partial_i\bm n)}d(\partial_i\bm n) \ , \
\label{pressure}
\end{eqnarray}
The relation (\ref{pressure}) is equivalent to the thermodynamic identity (\ref{wiakc4}).

\subsection{Non-dissipative equations}
\label{sec:reac}

The non-dissipative dynamic equations can be constructed using the Poisson brackets method, see Refs. \cite{DV80,KL93}. As it is known from classical mechanics \cite{LL78}, in terms of the Poisson brackets method dynamic equation for a variable $\phi$ is
\begin{eqnarray}
\label{k1}
\frac{\partial \phi }{\partial t} = \{{\cal H} , \phi \},
\end{eqnarray}
where ${\cal H}$ stands for Hamiltonian (energy of the system), and $\{...\}$ designate the Poisson brackets. Unlike classical mechanics, the hydrodynamic variables are fields, depending on both, time and space coordinates, and Hamiltonian is an integral over space of the energy density $U$ dependent of the fields. For the nematics
\begin{equation}
{\cal H} = \int dV \, U(\bm D_0, \bm B, \bm J, S, \rho, \bm n,\partial_i\bm n) .
\label{Ham}
\end{equation}
Nevertheless, the dynamic equations for the fields can be written in the form (\ref{k1}) by exploiting the proper expressions for their Poisson brackets. Since the macroscopic dynamic equations are local, the Poisson brackets for any pairs of hydrodynamic variables have to be proportional to delta-function (or its space derivative).

Following methods of Refs. \cite{DV80,KL93} we present the full set of the Poisson brackets needed to study nematodynamics in the presence of the electomagnetic field (including the electromagnetic field itself). The Poisson brackets for all pairs of the hydrodynamic variables are known \cite{DV80,KL93}, non-zero Poisson brackets are
\begin{eqnarray}
\left\{J_i(\bm r), \rho(\bm x) \right\}
= \rho(\bm r) \partial_i \delta(\bm r-\bm x) ,
\label{pois2} \\
\left\{J_i(\bm r), S(\bm x) \right\}
= S (\bm r)\, \partial_i \delta(\bm r-\bm x) ,
\label{pois3} \\
\left\{J_k(\bm r), J_i(\bm x) \right\}
= J_i(\bm r) \partial_k \delta(\bm r-\bm x)
+ \partial_i \delta(\bm r-\bm x) \, J_k(\bm x) .
\label{pois4}
\end{eqnarray}
Here derivatives like $\partial_i  \delta(\bm r-\bm x)$ mean $\partial  \delta(\bm r-\bm x)/\partial r_i$.

For the Poisson bracket $\{ J_i, n_k\}$, one can use the expression (\ref{unit1}) or (\ref{unit2}), derived for the vector fields. The expressions  correspond to ideal rods or discs frozen into the liquid crystal material \cite{Volovik1980}. For definiteness, we use the expression (\ref{unit1}), that is
\begin{equation}
\{ J_i(\bm r), n_k(\bm x)\}
=-\partial_i n_k \delta(\bm r-\bm x)
-\partial_j \delta(\bm r-\bm x) \delta^\perp_{ik}(\bm x) n_j(\bm x),
\label{pois1}
\end{equation}
where $\delta^\perp_{ik}=\delta_{ik}-n_i n_k$. The second case (\ref{unit2}) can be written similarly using the reactive kinetic term (see below).

In addition, one can derive the expression for Poisson bracket of the electromagnetic field
\begin{eqnarray}
\left\{ B_j(\bm r), D_{0k}(\bm x) \right\}
=4\pi c \,\epsilon_{jmk}\partial_m\delta(\bm r-\bm x),
\label{pois5}
\end{eqnarray}
see Appendix \ref{sec:lagrange}. Here $\epsilon_{jmk}$ is Levi-Civita antisymmetric tensor. The Poisson brackets of $\bm B,\bm D_0$ with other variables are zero. It implies, particularly, that we do not include the electromagnetic contribution to the momentum density $\bm J$. Exploiting the expression (\ref{pois5}), one immediately finds from Eq. (\ref{wiakc4}) the equations (\ref{maxw1},\ref{maxw2}), where $\bm D,\bm H$ are substituted by $\bm D_0,\bm H_0$:
\begin{equation}
\partial_t \bm D_0
=c \nabla\times \bm H_0.
\label{maxw4}
 \end{equation}
The same result is obtained in Appendix \ref{sec:lagrange}, see Eq. (\ref{lagr27}).

With the all Poisson brackets (\ref{pois2}-\ref{pois1}) in hands we obtain the full set of non-dissipative, nonlinear dynamic equations for nematics
\begin{eqnarray}
\partial_t \rho =\{{\cal H},\rho\}=-\nabla \bm J, \quad
\partial_t S =\{{\cal H},S\} = - \nabla(S\bm v),
\label{pois6} \\
\partial_t n_i = \{{\cal H},n_i\}  =
-\bm v \nabla n_i +n_j \delta^\perp_{ik} \partial_j v_k.
\label{pois7}
\end{eqnarray}
Remind that $\bm v=\bm J/\rho$. The equations (\ref{pois6},\ref{pois7}) have the same form as in absence of the electromagnetic field. The equation for the momentum density can be written as
\begin{eqnarray}
\partial_t J_i =\{{\cal H},J_i\}
=-\partial_k(v_k J_i)
-\nabla \left( \frac{\partial U}{\partial(\nabla n_j)}\partial_i n_j\right)
-\partial_k(\Xi_j \delta^\perp_{ij}n_k)-\partial_i p
\nonumber \\
+\frac{1}{4\pi}\partial_k \left(D_{0k} E_i+B_k H_{0i}\right)
-\frac{1}{4\pi c} \partial_t [\bm D_0\times \bm B]_i,
\label{momentum}
\end{eqnarray}
where
\begin{equation}
\Xi_i=-\frac{\partial U}{\partial n_i}
+\partial_k\frac{\partial U}{\partial (\partial_k n_i)}.
\label{dirder}
\end{equation}

The last term in Eq. (\ref{momentum}), derived, using Maxwell equations (\ref{maxw2},\ref{maxw3},\ref{maxw4}), is merely (with the sign minus) time derivative of the momentum density of the electromagnetic field, see Appendix \ref{sec:lagrange}. The equations (\ref{pois6}-\ref{momentum}) lead to the energy conservation law
\begin{eqnarray}
\partial_t U =  -\, \nabla \bm Q^{(r)} = -\frac{c}{4\pi}\,\nabla[\bm E\times \bm H]
-\nabla\left[\frac{\partial U}{\partial(\nabla n_i)}
(\bm v \nabla n_i - n_k \delta^\perp_{ij}\partial_k v_j)\right]
\nonumber \\
-\partial_i\left[(\rho\mu+TS+\bm J \bm v)v_i\right]
-\partial_k(\Xi_jn_k \delta^\perp_{ij}v_i).
\label{pois10}
\end{eqnarray}
where the non-dissipative part of energy flow density is designated  by $ \bm Q^{(r)}$.

We consider relatively small frequencies and non-relativistic hydrodynamic motions. In this situation the induced magnetic field is weak, one can say, that $H$ has the smallness $v/c$. However, the magnetic field cannot be excluded from the equations. The reason is that there is the contribution to the equation for $U$ (\ref{pois10}) related to the electromagnetic energy flow, its density is determined by Poynting vector
\begin{equation}
\bm S= \frac{c}{4\pi} \bm E \times \bm H.
\label{Poynting}
\end{equation}
Thus though the magnetic field has the smallness $v/c$, the expression for electromagnetic energy flow (\ref{Poynting}) contains the compensation factor $c$.

\subsection{Kinetic terms}
\label{diss}

Now it is necessary to add kinetic terms to the obtained dynamic equations (\ref{pois6}-\ref{pois10}).
The general scheme of constructing the terms based on the fact that these terms are proportional to the variational derivatives of the energy over the dynamic variables (or their  derivatives) with some coefficients (kinetic coefficients). The kinetic coefficients have to be chosen to satisfy Onsager symmetry and to lead to positive entropy production. Presented below kinetic contributions into dynamic equations can be found separately in the literature \cite{GP93,KL93,PB96,LL10,TC16,KP08,EC19}. We present these terms because to our knowledge they never have been concisely derived altogether (for hydrodynamic, director, and electromagnetic degrees of freedom). Besides in what follows we discuss shortly physical meanings of the entering equations terms. We do believe it leads to a deeper insight into the mathematical structures of the equations, what may be helpful to the readers.

In this way the equations including dissipative terms can be found by adding of the following dissipative
contributions to equations (\ref{pois6}-\ref{pois10})  and (\ref{maxw4})
\begin{eqnarray}
 \partial_t  n_i = \{{\cal H}, n_i \}  +  \frac{1}{\gamma} \Xi_i
-\frac{1-\lambda}{2}(n_j \delta^\perp_{ik}+n_k \delta^\perp_{ij}) \partial_kv_j,
\label{diss1} \\
 \partial_t D_{0i} =
c \epsilon_{ikn}\partial_k H_n
-  4\pi \sigma_{ik}E_k \, ,
\label{diss2} \\
 \partial_t J_i =  \{{\cal H},J_i\}+ \partial_k(\eta_{iknm}\partial_n v_m)
+\partial_k \left[\frac{1-\lambda}{2}(n_i \delta^\perp_{jk}+n_k \delta^\perp_{ij})\Xi_{j}\right],
\label{diss3} \\
 \partial_t S = \{{\cal H},S\} +
\partial_i\left(\frac{\kappa_{ik}}{T} \partial_k T\right)+\frac{R}{T} \, ,
\label{entropy} \\
 \partial_t U = -\, \nabla \bm Q^{(r)} +  \partial_i(\kappa_{ik} \partial_k T)
+\nabla\left(\frac{\partial U}{\partial(\nabla n_i)}\frac{\Xi_i}{\gamma}\right)
+\partial_i(v_k \eta_{iknm}\partial_n v_m)
\nonumber \\
-\nabla\left[\frac{\partial U}{\partial(\nabla n_i)}
\frac{1-\lambda}{2}(n_j \delta^\perp_{ik}+n_k \delta^\perp_{ij}) \partial_kv_j\right]
+\partial_k\left[v_i \frac{1-\lambda}{2}(n_i \delta^\perp_{jk}+n_k \delta^\perp_{ij})\Xi_{j}\right],
\label{diss5} \\
\end{eqnarray}
where
\begin{equation}
R=\frac{\kappa_{ik}}{T}\partial_i T\partial_k T
+\frac{1}{\gamma } \Xi^2
+\eta_{iknm}\partial_i v_k \partial_n v_m
+\sigma_{ik}E_i E_k.
\label{diss4}
\end{equation}
Here $\sigma_{ik}$ is the electric conductivity tensor, $\kappa_{ik}$ is the thermal conductivity tensor, $\eta_{iknm}$ is the viscosity tensor, $\gamma$ is the coefficient of the rotational viscosity and $\lambda$ is some kinetic coefficient.

Onsager symmetry leads to the conclusions, that the tensors $\sigma_{ik}$ and $\kappa_{ik}$ are symmetric
and that $\eta_{iknm}=\eta_{nmik}$. The kinetic contribution to the stress tensor should be symmetric.
The term with the kinetic coefficient $\lambda$ is explicitly symmetric. The symmetry of the viscous contribution implies $\eta_{iknm}=\eta_{kinm}$. Since $R/T$ is the entropy production rate, $R$ should be positive. It imposes some inequalities on the components of the tensors $\sigma_{ik}$, $\kappa_{ik}$, $\eta_{iknm}$ and leads to $\gamma>0$. The important point should be made here is that the terms with the kinetic coefficient $\lambda$ do not contribute to the entropy production rate, such kinetic terms are called reactive. Note that for $\lambda=-1$ the sum of the reactive terms and of the terms obtained via the Poisson brackets, is equivalent to using the alternative expression for the Poisson bracket $\{J_i,n_k\}$ that can be obtained from Eq. (\ref{unit2}). It corresponds to the case of ideal disc molecules frozen in the liquid crystal material \cite{Volovik1980}.

The equation (\ref{diss2}) has to be compared with the Maxwell equation (\ref{maxw1}). We conclude that
\begin{equation}
\partial_t D_i=\partial_t D_{0i}
+4\pi \sigma_{ik}E_k.
\label{dissad}
\end{equation}
The last term in Eq. (\ref{dissad}) is a generalization of the usual contribution to the displacement field
related to conductivity \cite{LL84}, for the case of the anisotropic medium (nematic). Above we ignored a dissipative contribution to the magnetizing field $\bm H$. It is justified if we consider the case $v\ll c$.
In the case where the magnetic field $\bm B$ is much larger than the electric field $\bm E$,
corrections related to the nematic velocity should be taken into account. The case is analyzed in Appendix \ref{sec:lorentz}.

In the nematic phase the tensors $\sigma_{ik}$, $\kappa_{ik}$, $\eta_{iknm}$ can be written in terms of the director $\bm n$. Say,
\begin{eqnarray}
\sigma_{ik}=\sigma_\parallel n_i n_k
+\sigma_\perp (\delta_{ik}-n_i n_k),
\nonumber \\
\kappa_{ik}=\kappa_\parallel n_i n_k
+\kappa_\perp (\delta_{ik}-n_i n_k).
\nonumber
\end{eqnarray}
The forth rank viscosity tensor $\eta_{iknm}$ is generally determined by five independent viscosity coefficients, according to five independent combinations constructed by the unit Kronecker's delta tensor $\delta_{ik}$, and the director ${\bm n}$. One can easily write down, e.g., the following forms \cite{LL86}:
\begin{itemize}
\item
$n_i n_k n_n n_m$
\item
$n_i n_k\delta _{n m} + n_i n_n \delta _{m k}$
\item
$n_i n_n\delta _{km} + n_k n_n \delta _{im} + n_i n_m \delta _{kn} + n_k n_m \delta_{in}$
\item
$\delta _{ik} \delta _{nm}$
\item
$\delta _{in}\delta _{km} + \delta _{kn}\delta _{im}$
\end{itemize}
Unfortunately in the literature on liquid crystals there is still no unified notation for the coefficients in front of these combinations (as well as the choice these combinations themselves). We follow the notation adopted in the classic textbook on liquid crystals \cite{GP93}, and then the viscous part of the stress tensor, reads as
\begin{eqnarray}
\tau _{ij}^{(v)} \equiv \eta_{ijkl} A_{kl} = 2\nu _2 A_{ij} + 2(\nu_3 - \nu_2)(A_{ik}n_kn_j + A_{jk}n_i n_k)
\nonumber \\
\label{new1}
+(\nu_4 - \nu _2)\delta_{ij}A_{kk} +2(\nu_1 + \nu_2 - 2\nu_3) n_in_jn_kn_lA_{kl} +
(\nu_5 - \nu_4 + \nu _2)(\delta _{ij}n_k n_lA_{kl} + n_i n_j A_{kk}),
\end{eqnarray}
where $\nu _1\div \nu_5$ - kinematic viscosity coefficients, and $A_{ij} \equiv (\partial _i v_j + \partial _j v_i)$. Then the entropy production related to the velocity is determined by $R^{(v)} = A_{ij}\,\tau _{ij}^{(v)}/2$.

Positiveness of the full entropy production, that is $R>0$, see Eq. (\ref{diss4}),  leads to the conditions
\begin{equation*}
\gamma>0, \ \sigma_\parallel>0, \
\sigma_\perp>0, \ \kappa_\parallel>0, \
\kappa_\perp>0.
\end{equation*}
The conditions for the viscosity tensor $\eta_{iknm}$ can be formulated in terms of the viscosity coefficients (\ref{new1}).
The entering (\ref{new1}) coefficients should satisfy the following conditions
\begin{eqnarray}
\label{new2}
\nu_4 (2\nu_1 + \nu_2) > (\nu_5 -\nu _4)^2\, ;\,
\nu _2, \nu_3, \nu_4 > 0\, ;\, 2(\nu _1 + \nu _5) - \nu _4 + \nu _2 >0.
\end{eqnarray}
At time and space scales of typical liquid crystal devices and known material parameters of liquid crystals
(film thickness of an order of few micrometers, and the like) one can neglect of liquid crystalline material compressibility. For the viscous part of the stress tensor (\ref{new1}) it means that the terms proportional to $\nabla {\bm v}$ with the coefficients $\nu _5 $, and $\nu _2 - \nu _4$ can be neglected. Then the positivity of the entropy production (\ref{new2}) is reduced to the conditions presented in \cite{GP93}
\begin{eqnarray}
\label{new3}
\nu _1\, ; \, \nu _2\, ;\, \nu _3\, >\, 0 \ .
\end{eqnarray}

\section{Minimal model for nematics}
\label{sec:model}

Further we ignore effects related to magnetic polarization of the nematics. However there is another magnetic effect coming from a finite conductivity of the nematics. It is known that the electric current density of a moving conductor is equal to
\begin{equation}
j_i=\sigma_{ik} \left(E_i + \epsilon_{ikn}\frac{v_k}{c}B_n\right),
\nonumber
\end{equation}
see Ref. \cite{LL84}. Of course any achievable velocities of the nematics are much smaller than $c$.
However, the second contribution to the electric current density (related to the magnetic field) can be relevant if $B\gg E$. The case is analyzed in Appendix \ref{sec:lorentz}. Below we neglect the magnetic contribution to the current density.

In nematics the displacement field $\bm D_0$ in the main approximation can be written as
\begin{equation}
\bm D_0=\hat \epsilon \bm E + 4\pi \bm P_{fl}.
\label{wiakc1}
\end{equation}
Here $\bm E$ is electric field, the matrix $\hat \epsilon$ is the permittivity matrix of the nematic
and $\bm P_{fl}$ represents the flexoelectric contribution to the polarization vector \cite{Meyer69,Pikin1991}
related to a non-homogeneity of the director $\bm n$. The permittivity matrix and $\bm P_{fl}$ are
\begin{eqnarray}
\epsilon_{ik}=\epsilon_\perp(\delta_{ik}-n_in_k)
+\epsilon_\parallel n_i n_k,
\label{permit} \\
\bm P_{fl}= \zeta _1 \bm n (\nabla \bm n) + \zeta _2 (\bm n \nabla )\bm n ,
\label{weakc7}
\end{eqnarray}
where $\epsilon_\perp,\epsilon_\parallel,\zeta_1,\zeta_2$ are phenomenological coefficients,
characterizing dielectric permeability and flexoelectric response.

The energy $U$ in the same approximation is
\begin{eqnarray}
U= \frac{1}{8\pi}\bm D_0 \hat\epsilon^{-1} \bm D_0
-\bm P_{fl} \hat\epsilon^{-1}\bm D_0
+\frac{1}{2\rho}J^2 +F_F +U_0(S,\rho).
\label{poten2}
\end{eqnarray}
Then the identity $\bm E=4\pi \partial U/\partial \bm D_0$, see Eq. (\ref{wiakc4}),
reproduces Eq. (\ref{wiakc1}). The term ${F}_F$ in Eq. (\ref{poten2}) is Frank energy:
\begin{eqnarray}
F_F=\frac{K_1}{2}(\nabla \bm n)^2
+\frac{K_2}{2}[ \bm n (\nabla\times \bm n)]^2
+\frac{K_3}{2}[ \bm n \times (\nabla\times \bm n)]^2.
\label{Franken}
\end{eqnarray}
where $K_1,K_2,K_3$ are splay, twist and bend Frank modules.

Having in mind some specific and realistic nonlinear phenomena in liquid crystals, the formulated above equations can be simplified. For any realistic flows, the liquid crystal (nematic or smectic $A$) can be treated as incompressible, that is mass density $\rho$ is constant and $\nabla \bm v=0$. Furthermore, in the case of large thermal conductivity, temperature $T$ is homogeneous. In the opposite case of small thermal conductivity the specific entropy $S/\rho$ is homogeneous. Both limit cases enable one to exclude temperature (entropy) from the consideration. The set of equations can be made more compact assuming a single constant approximations for Frank elastic energy, and keeping only a single flexoelectric coefficient
(all the more that in the main approximation it is also the case for smectic $A$ liquid crystals, see Section \ref{sec:smec}).

In spite of these rigorously speaking erroneous assumptions, our model correctly identifies the important excitation modes, and their characteristic time and space scales. If necessary, these assumptions of the simplified model can be easily relaxed at the cost of more cumbersome set of equations. As a note of caution it is important not to overplay with such simplifications. Special care should be taken assuming a single (isotropic) viscosity coefficient. For example, for various types of electro-hydrodynamic instabilities in nematics, just the interplay between different viscosity coefficients determines the threshold of the instability. Another striking example provided  lyotropic smectic $A$ liquid crystal. In this case \cite{BF75} a simple shear within water layers is determined by the very low $10^{-2}\, Poise$  viscosity, while all other hydrodynamic motions include several hundred times larger membrane viscosity.

\section{Smectics}
\label{sec:smec}

To span a wide range of possibilities to apply our results, in this section we consider the nonlinear dynamics of a smectic-$A$. Instead of the director $\bm n$ smectic $A$ is characterized by the displacement $u$ of the smectic layers in $z$-direction, where $z$-axis is perpendicular to the equilibrium positions of the smectic layers. It is convenient to formulate the equations in terms of the variable $W=z-u$, what allows to formulate the equations in the invariant under rotations form (see e.g., \cite{GP93,KL93}). Then we deal with the thermodynamic identity
\begin{equation}
dU=\frac{1}{4\pi}\bm E d\bm D_0+\frac{1}{4\pi}\bm H_0 d \bm B
+Td S+\mu d \rho + \bm v d \bm J
+\frac{\partial U}{\partial (\partial_i W)}d(\partial_i W)
+M_{ik}d(\partial_i\partial_k W),
\label{smec1}
\end{equation}
instead of Eq. (\ref{wiakc4}).

In the main approximation the smectic elastic energy contributions into $U$ can be written as
\begin{equation}
U_{sm}=\frac{B}{8}[(\nabla W)^2-1]^2+\frac{K}{2} (\nabla^2 W)^2,
\label{smec2}
\end{equation}
instead of Eq. (\ref{Franken}). Therefore
\begin{eqnarray}
U= \frac{1}{8\pi}\bm D_0 \hat\epsilon^{-1} \bm D_0
-\bm P_{fl} \hat\epsilon^{-1}\bm D_0
+\frac{1}{2\rho}J^2 +U_{sm} +U_0(S,\rho).
\nonumber
\end{eqnarray}
The permittivity matrix $\hat \epsilon$ and the flexoelectric contribution to the polarization vector
$\bm P_{fl}$ of the smectic are written as
\begin{eqnarray}
\epsilon_{ik}=\epsilon_\perp(\delta_{ik}-l_il_k)
+\epsilon_\parallel l_i l_k,
\label{permit2} \\
\bm P_{fl}= \zeta _1 \bm l (\nabla \bm l) + \zeta _2 (\bm l \nabla )\bm l ,
\label{smec3}
\end{eqnarray}
instead of Eqs. (\ref{permit},\ref{weakc7}). In the expressions (\ref{permit2},\ref{smec3}) $\bm l$
is the unit vector perpendicular to the smectic layers: $\bm l= \nabla W/|\nabla W|$.
It is worth to note, that pure flexoelectric instability in smectic $A$ liquid crystals
(as far as we know) was not observed experimentally. However there is no doubts about the very existence of the flexoelectric effect
in smectics $A$ (see e.g., \cite{PP76}, where the coefficient $\zeta _1$ has been measured).

To formulate the non-dissipative dynamic equations of the smectics we use the same Poisson bracket method. We should use the Poisson bracket
\begin{equation}
\{\bm J(\bm r),W(\bm x)\}=
-\nabla W \delta(\bm r-\bm x),
\label{smec4}
\end{equation}
instead of Eq. (\ref{pois1}). The other expressions (\ref{pois2}-\ref{pois5}) for the Poisson brackets remain unchanged. Then the equation for the variable $W$ is
\begin{equation}
\partial_t W=\{H,W\}=-\bm v \nabla W, \quad
\partial_t u=v_z-\bm v \nabla u,
\label{smec5}
\end{equation}
where we used the expressions (\ref{smec1},\ref{smec4}). The equation for the displacement $u$ is obtained after the substitution $W=z-u$.

The equation for the momentum density of the smectic is
\begin{eqnarray}
\partial_t J_i =
-\partial_k(v_k J_i)
-\partial_k \left( M_k\partial_i W\right)
-\partial_k(M_{kj}\partial_i\partial_j W)
\nonumber \\
-\partial_i p
+\frac{1}{4\pi}\partial_k \left(D_{0 k} E_i+B_k H_{0 i}\right) - \frac{1}{4\pi c} \partial_t [\bm D_0\times \bm B]_i  \,  ,
\label{smec6}
\end{eqnarray}
instead of Eq. (\ref{momentum}). Here
\begin{equation}
M_i=\frac{\partial U}{\partial(\partial_i W)}
-\partial_k M_{ik} \ , \
\nonumber
\end{equation}
and
\begin{equation}
M_{ik}=\frac{\partial U}{\partial(\partial_i \partial_k W)}
\ . \
\nonumber
\end{equation}
Pressure $p$ is determined by the same relation (\ref{pressure2}).

One should add dissipative terms to the written above reactive equations for the smectics.
The dissipative terms in the equations for electric displacement field and momentum density
are analogous to the dissipative terms in equations (\ref{diss2},\ref{diss3}) where one should substitute $\bm n \to \bm l$.
The dissipative contributions to the equations for $W$ and energy density $U$ are
\begin{eqnarray}
&&\partial_t W=-\bm v \nabla W +\Theta, \quad
\Theta= -\xi_1 |\nabla W|^2 h
-\frac{\xi_2}{T}\nabla W \nabla T,
\label{diss43} \\
&&\partial_t U =   -\, \nabla \bm Q^{(r)}_{Sm}  + \partial_i(\kappa_{ik} \partial_k T)
+\partial_i(v_k \eta_{iknm}\partial_n v_m)
\nonumber \\
&&+\nabla\left(\frac{\partial U}{\partial(\nabla W)}\Theta\right)
+\partial_i (M_{ik}\partial_k \Theta)
-\partial_k (\partial_i M_{ik}\Theta),
\label{diss55}
\end{eqnarray}
where
\begin{eqnarray}
h=-\nabla \frac{\partial U}{\partial(\nabla W)}
+\partial_i\partial_k M_{ik},
\label{smec8}
\end{eqnarray}
is the variational derivative of the energy over $W$, in turn non-dissipative energy flow  density for smectics is
\begin{eqnarray}
 \big(Q^{(r)}_{Sm}\big)_i = \frac{c}{4\pi}\,[\bm E\times \bm H]_i
+   M_i  \,(v_k \partial_k W) + M_{ik}  \,\partial_k (v_l \partial_l W)
\nonumber \\
+\,  \left(\rho\mu+TS+\bm J \bm v \right) v_i
\ . \
\label{pois10}
\end{eqnarray}
In Eq. (\ref{diss43}) $\xi_1,\xi_2$ are the permeation coefficients. The equation for entropy density is
\begin{eqnarray}
&&\partial_t  S= \{ S,H\} +
\partial_i\left(\frac{\kappa_{ik}}{T} \partial_k T
+\frac{\xi_2}{T} \partial_iW h \right)+\frac{R}{T},
\label{entropy5} \\
&&R=\frac{\kappa_{ik}}{T}\partial_i T\partial_k T
+\xi_1 (\nabla W)^2 h^2
+2 \frac{\xi_2}{T} h \,(\nabla W)\, \nabla T
+\eta_{iknm}\partial_i v_k \partial_n v_m
+\sigma_{ik}E_i E_k.
\label{diss45}
\end{eqnarray}
The positive entropy production implies $\xi_1>0$, $\kappa_\parallel \xi_1>\xi_2^2/T$,
in addition to the conditions, analogous to ones formulated for nematics.

\section{Conclusions}
\label{sec:con}

To conclude we would like to stress that liquid crystals are far from being exhausted as a topic of research. Since typically liquid crystals are soft (easily excited and deformed) systems,
nonlinear physics is one of the prominent direction to study liquid crystals and their applications. In this work we have shown how to include electromagnetic field in the description of nonlinear dynamic phenomena in nematic and smectic $A$ liquid crystals in external a.c. electric field. Our main result (the formulated set of dynamic equations for nematic and smectic $A$ liquid crystals) does allow an entry point to study different nonlinear dynamic phenomena in liquid crystals. To name a few
\begin{itemize}
\item
Already mentioned electrically driven dynamic three-dimensional localized and moving excitations (directrons)
\item
Classical driven by electric field coarsening kink (separating stable and unstable director configurations) dynamics. Several interesting questions, requiring the complete set of nonlinear dynamic equations, the   naturally arise here.
\item
Dynamic transitions between isotropic and nematic liquid crystal phases in nonequilibrium a.c. driven systems (see, e.g., \cite{HU22}).
\item
Anisotropic viscous flow effects created in liquid crystals by rotating colloidal particles (see said above about assumption on a single isotropic shear viscosity coefficients, and recent publication \cite{LL22}).
\item
Dynamic of defects (disclinations in director field and hydrodynamic vortices in a driven nematic liquid crystal cell, see \cite{CF22}).
\end{itemize}
In all cases enlisted above the instability is triggered either by external electric field (which influences the system via dielectric anisotropy, flexoelectric coupling, or (in smectics) via so-called electroclinic effect \cite{KH07}, \cite{BL11}) or by externally imposed flow. Then well above the instability thresholds the system behavior controlled by nonlinear phenomena.

It is worth to add also that derived in this work nonlinear dynamic equations are especially important for chemically or electrically
driven active and biological orientationally ordered systems (see \cite{ZM21}). Indeed, equilibrium (static) atomic or molecular positions,
are essential for a dead structure. Biological functions are associated with molecular motions. All these studies require rather involved numeric
solutions of the nonlinear set of differential equations. This is beyond the scope of our work, which focuses only on analytic
derivation of the dynamic equations for nematic and smectic $A$ liquid crystals.

\acknowledgements

The work of E.I.K. and  E.S.P. was supported by the State assignment N. 0029-2019-0003,
V.V.L. thanks the support of the Russian Ministry of Science and Higher Education, Project No. 075-15-2022-1099.
The work of A.R.M. was supported by the State Assignment   FMME-2022-0008  (N. 122022800364-6).

\appendix

\section{Derivation of Poisson brackets}
\label{sec:lagrange}

Here we derive some expressions for Poisson brackets needed to construct the nonlinear equations presented in the paper. Note that the expressions are universal \cite{DV80,KL93} that is are independent of the concrete form of Hamiltonian (energy of the system).

\subsection{Vector fields}

Let us consider Hamiltonian dynamics for a system described by canonically conjugated vector fields $\bm p,\bm q$. Hamiltonian of the system is written as
\begin{equation}
{\cal H}= \int dV\, H(\bm p, \bm q, \partial_i \bm q).
\label{lagr6}
\end{equation}
Canonical equations for the fields $\bm p,\bm q$ are written as
\begin{eqnarray}
\partial_t \bm p= -\frac{\delta{\cal H}}{\delta q}
=-\frac{\partial H}{\partial q}+\partial_i \frac{\partial H}{\partial(\partial_i \bm q)},
\label{lagr7} \\
\partial_t \bm q =\frac{\delta{\cal H}}{\delta p}=\frac{\partial H}{\partial p}.
\label{lagr8}
\end{eqnarray}
The equations can be rewritten as
\begin{equation}
\partial_t \bm p =\{{\cal H},\bm p\}, \quad
\partial_t \bm q=\{{\cal H},\bm q\},
\label{cpois}
\end{equation}
where $\{\dots,\dots\}$ designate Poisson brackets. Non-zero Poisson brackets for the canonically conjugated fields $\bm p,\bm q$ are
\begin{equation}
\{ p_i(\bm r),q_k(\bm x)\}
=\delta_{ik}\delta(\bm r-\bm x).
\label{canon}
\end{equation}
Being substituted into Eq. (\ref{cpois}), the expression (\ref{canon}) leads to Eqs. (\ref{lagr7},\ref{lagr8}).

The equations (\ref{lagr7},\ref{lagr8}) lead to the following equation for the canonical momentum density
\begin{eqnarray}
\partial_t(-\bm p\partial_i \bm q)+\partial_k \Pi_{ik}=0,
\label{lagr9} \\
\Pi_{ik}=\left(\bm p \frac{\partial H}{\partial \bm p}-H\right)\delta_{ik}
+\frac{\partial H}{\partial(\partial_k \bm q)}\partial_i \bm q.
\label{lagr10}
\end{eqnarray}
However, the stress tensor (\ref{lagr10}) is not symmetric. Therefore the conservation of the angular momentum law based on the canonical momentum density is not granted. To overcome the difficulty we exploit rotational invariance of $H$
\begin{equation}
\epsilon_{ikn}\left[
p_k \frac{\partial H}{\partial p_n}
+q_k \frac{\partial H}{\partial q_n}
+\partial_j q_k \frac{\partial H}{\partial (\partial_j q_n)}
+\partial_k q_j \frac{\partial H}{\partial (\partial_n q_j)}
\right]=0,
\nonumber
\end{equation}
we obtain
\begin{equation}
\epsilon_{ikn}\Pi_{kn}
=-\epsilon_{ikn}\partial_t(p_k q_n)
-\epsilon_{ikn} \partial_j\left[q_k \frac{\partial H}{\partial (\partial_j q_n)}\right].
\label{lagr11}
\end{equation}
Using (\ref{lagr11}) and the standard relation $\epsilon_{ikn}\epsilon_{njl} = \delta_{ij}\delta_{kl} - \delta_{il}\delta_{kj}$
the equation (\ref{lagr9}) can be rewritten as
\begin{eqnarray}
&&\partial_t(-\bm p\partial_i \bm q)+\frac{1}{2}\partial_k (\Pi_{ik}+\Pi_{ki})
+\frac{1}{2}\partial_k (\Pi_{ik}-\Pi_{ki})=0 \ ,
\label{lagr101}
\end{eqnarray}
 where
\begin{eqnarray}
&&\partial_k (\Pi_{ik}-\Pi_{ki})=
\partial_k (\epsilon_{ikn}\epsilon_{njl}\Pi_{jl})=
-\partial_t \partial_k (p_i q_k-p_k q_i)-\epsilon_{ikn}\epsilon_{njl}\partial_k
\partial_m\left[ q_j \frac{\partial H}{\partial (\partial_m q_l)}\right] \ . \
\label{lagr12}
\end{eqnarray}
The last term in Eq. (\ref{lagr12}) can always be rewritten as a derivative of a symmetric tensor, using the identity
\begin{eqnarray}
\partial_k \partial_j M_{kji}=\partial_k \partial_j \left[M_{kji}
-\frac{1}{2}\epsilon_{kji}(\epsilon_{pqs}M_{pqs})
-\epsilon_{kjs}\epsilon_{qls}M_{qil}\right].
\label{symmet}
\end{eqnarray}
One can easily check, that the combination in the square brackets in Eq. (\ref{symmet}) is invariant under the permutation $i\leftrightarrow k$. Therefore the equation (\ref{lagr101}) can be rewritten as
\begin{eqnarray}
\partial_t\left[-\bm p\partial_i \bm q+\partial_k(p_k q_i)\right]
-\partial_k T_{ik}=0,
\label{lagr13}
\end{eqnarray}
where the stress tensor $T_{ik}$ is symmetric.

Thus we conclude that for vector fields the correct form for the momentum density is
\begin{equation}
J_i=-\bm p \partial_i \bm q +\partial_k(p_k q_i).
\label{lagr14}
\end{equation}
Besides the canonical contribution, the expression (\ref{lagr14}) contains the additional term characteristic of the vector fields. Using the relation (\ref{canon}), one can derive from Eq. (\ref{lagr14}) the expression
\begin{equation}
\left\{J_k(\bm r), J_i(\bm x) \right\}
=J_i(\bm r) \partial_k \delta(\bm r-\bm x)
+\partial_i \delta(\bm r-\bm x) J_k(\bm x),
\nonumber
\end{equation}
coinciding with Eq. (\ref{pois4}), thus confirming its universality.

Starting with the relation (\ref{canon}), one derives from the expression (\ref{lagr14}) the following expressions for the Poisson brackets
\begin{eqnarray}
\left\{ J_i(\bm r), p_k(\bm x)\right\}
=p_k(\bm r) \partial_i\delta(\bm r-\bm x)
-\delta_{ik} \partial_n \delta(\bm r-\bm x) p_n(\bm x),
\label{lagr15} \\
\left\{ J_i(\bm r), q_k(\bm x)\right\}
=-\partial_i q_k\delta(\bm r-\bm x)
+\partial_k \delta(\bm r-\bm x) q_i(\bm x),
\label{lagr16} \\
\left\{ J_i(\bm r), b_k(\bm x)\right\}
=b_k(\bm r) \partial_i \delta(\bm r-\bm x)
-\delta_{ik}\partial_j \delta(\bm r-\bm x) b_j(\bm x),
\label{lagr17}
\end{eqnarray}
where $\bm b= \nabla\times \bm q$. One derives from Eqs. (\ref{lagr15},\ref{lagr16}) for unit vectors
\begin{eqnarray}
\left\{ J_i(\bm r), \frac{p_k}{p}(\bm x)\right\}
= - \Big(\partial_i \frac{p_k}{p}\Big)\, \delta(\bm r-\bm x)
-\partial_n \delta(\bm r-\bm x)
\left[ \frac{p_n}{p}\left(\delta_{ik}-\frac{p_i}{p}\frac{p_k}{p}\right)\right],
\label{unit1} \\
\left\{ J_i(\bm r), \frac{q_k}{q}(\bm x)\right\}
= - \Big(\partial_i \frac{q_k}{q}\Big) \, \delta(\bm r-\bm x)
+\partial_j  \delta(\bm r-\bm x)
\left[\left(\delta_{jk}-\frac{q_j}{q}\frac{q_k}{q}\right) \frac{q_i}{q}\right],
\label{unit2}
\end{eqnarray}
where the expressions in square brackets are functions of $\bm x$.

\subsection{Electromagnetic field}

We begin with non-dissipative dynamic equations for the electromagnetic field in Lagrangian formulation. It is convenient to exploit Weyl gauge where both the electric field $\bm E$ and the magnetic field $\bm B$ are expressed in terms of the vector potential $\bm A$ as
 \begin{equation}
 \bm E=-\frac{1}{c}\partial_t \bm A, \quad
 \bm B=\nabla \times \bm A.
 \label{lagr1}
 \end{equation}
The density of Lagrange function $L$ depends on $\bm E, \bm B$, that is on the derivatives of the vector potential $\bm A$,
in the agreement with Eq. (\ref{lagr1}). This density $L$ is related to the density of the internal energy $U$ via Legendre transform
\begin{equation}
U=\frac{1}{4\pi}\bm D_0 \bm E -L, \quad
\bm D_0 = 4\pi \frac{\partial L}{\partial \bm E}.
\label{lagr2}
 \end{equation}
Thus Hamiltonian (energy of the system) is written as
\begin{eqnarray}
{\cal H}=\int dV\, U(\bm D_0,\bm B,\dots),
\label{lagr29} \\
dU=\frac{1}{4\pi} \bm E d \bm D_0 +\frac{1}{4\pi} \bm H_0 d \bm B +\dots ,
\label{lagr30}
\end{eqnarray}
where dots stand for other variables.

One obtains from Eqs (\ref{lagr2},\ref{lagr30})
\begin{equation}
dL =\frac{1}{4\pi} \bm D_0 d \bm E -\frac{1}{4\pi} \bm H_0 d \bm B +\dots .
\label{lagr3}
\end{equation}
Therefore the variable, canonically conjugated to $\bm A$, is
\begin{equation}
\frac{\partial L}{\partial(\partial_t \bm A)}
=-\frac{\bm D_0}{4\pi c}.
\label{lagr4}
\end{equation}
Thus, we arrive at the following expression for the Poisson bracket
\begin{equation}
\left\{A_i(\bm r), D_{0 k} (\bm x) \right\}
=4\pi c \delta_{ik} \delta(\bm r-\bm x).
 \label{lagr5}
 \end{equation}
Taking curl of the relation (\ref{lagr5}), we find
\begin{equation}
\left\{B_j(\bm r), D_{0 k}(\bm x) \right\}
=4\pi c \,\epsilon_{jmk}\partial_m\delta(\bm r-\bm x) ,
\label{lagr26}
\end{equation}
that is at Eq. (\ref{pois5}).

Now we can derive the canonical equations for the electromagnetic field. Using the thermodynamic identity (\ref{lagr2})
and the expression (\ref{lagr26}), we end up with Maxwell equations
\begin{eqnarray}
\partial_t \bm D_0=\{ {\cal H}, \bm D_0 \}
=c \nabla \times \bm H_0,
\label{lagr27} \\
\partial_t \bm B=\{ {\cal H}, \bm B \}
=-c \nabla \times \bm E.
\label{lagr28}
\end{eqnarray}
Note that the equation (\ref{lagr28}) is a direct consequence of the relations (\ref{lagr1}). One can find the contribution to the momentum density related to the electromagnetic field. Substituting to the expression (\ref{lagr14}) $\bm p=-(4\pi c)^{-1}\bm D_0$, $\bm q=\bm A$,
one finds
\begin{equation}
\bm J^{em}=\frac{1}{4\pi c} \bm D_0\times \bm B,
\label{lagr31}
\end{equation}
where we assumed $\nabla \bm D_0=0$. The condition means zero density of free charges.

\section{Conductivity at finite velocity}
\label{sec:lorentz}

Here we pass to the thermodynamic potential.
\begin{eqnarray}
\tilde U=U- \frac{1}{4\pi} \bm B \bm H_0,
\label{lor3} \\
d\tilde U= \frac{1}{4\pi} \bm E\, d \bm D_0
-\frac{1}{4\pi} \bm B\, d \bm H_0 +\dots.
\label{lor4}
\end{eqnarray}
Thus, the time derivative of the energy density is
\begin{equation}
\partial_t \tilde U=
 \frac{1}{4\pi} \bm E\, \partial_t \bm D_0
-\frac{1}{4\pi} \bm B\, \partial_t \bm H_0 +\dots,
\label{lor5}
 \end{equation}
where we present the electromagnetic terms. We are interested in the dissipative contribution to $\partial_t \tilde U$.

If the velocity $\bm v$ of a system  is nonzero then the relation between the current density of the free charges and the electric field $\bm E$ is
\begin{equation}
j_i  =  \sigma_{ik}\left(E_k+\epsilon_{knm}\frac{v_n}{c}B_m\right).
\label{lor1}
\end{equation}
It is a consequence of the fact, that the vectors $\bm E, \bm B$ and the vectors $\bm D,\bm H$
can be presented as the components of four-dimensional antisymmetric tensors of second order (with
the condition that the charge density is zero) \cite{LL84}. The arguments behind Eq. (\ref{lor1}) are based on Lorentz invariance, and go back to Minkowski (1907). The quantity (\ref{lor1})  is the kinetic (dissipative) contribution to $-\partial_t \bm D_0/(4\pi)$. It is interesting to note that the second contribution in Eq. (\ref{lor1}) can be relevant even at $v/c\ll1$ if $B\gg E$. Similar arguments lead to the conclusion that the dissipative contribution to the magnetization derivative $\partial_t H_{0i}/(4\pi)$ is
\begin{equation}
\epsilon_{ilj}\frac{v_l}{c}
\sigma_{jk}\left(E_k+\epsilon_{knm}\frac{v_n}{c}B_m\right).
\label{lor2}
\end{equation}
We find from Eqs. (\ref{lor4},\ref{lor1},\ref{lor2}) that
\begin{equation}
R =   \frac{1}{4\pi}\sigma_{ij}\left(E_i+\epsilon_{ink}{\frac{v_n}{c}B_k}\right)
\left(E_j+\epsilon_{jml}{\frac{v_m}{c}B_l}\right) +\dots,
\label{lor6}
\end{equation}
where $R/T$ is the entropy production rate. Thus, the only difference in comparison with the scheme presented in the main text is in the substitution $\bm E\to \bm E'$ where
\begin{equation}
\bm E' =\bm E+
\frac{\bm v}{c}\times \bm B \ , \
\label{lor7}
\end{equation}
(compare with \cite{LL84}, Eq. (63.1)).
Let us stress that the relations derived in the section, are correct at an arbitrary velocity $\bm v$.

\end{document}